\documentclass[12]{article}
\usepackage{amssymb}
\usepackage{a4}
\evensidemargin \oddsidemargin
\newcommand{\beq}{\begin{equation}}
\newcommand{\eeq}{\end{equation}}
\newcommand{\ba}{\begin{array}}
\newcommand{\ea}{\end{array}}

\newcommand{\sect}[1]{\setcounter{equation}{0}\section{#1}}

\newcommand{\bea}{\begin{eqnarray}}
\newcommand{\eea}{\end{eqnarray}}
 
\newtheorem{theorem}{Theorem}[section]

\newtheorem{naming}{Definition}[section]   
\newcommand{\bdefi}{\medskip\begin{naming} ~~\\ \it}
\newcommand{\edefi}{\end{naming} \hfill $\Diamond$ }
\def\sq{\mbox{\rlap{$\sqcap$}$\sqcup$}}
\newenvironment{proof}[1]{\vspace{5pt}\noindent{\sc Proof #1}\hspace{6pt}}%
{\hfill\sq}
\newcommand{\bp}{\begin{proof}}
\newcommand{\ep}{\end{proof}\par\vspace{10pt}\noindent}
\def\b#1{{\mathbb #1}}

\def\nn{\nonumber  \\}

\date{}

\begin{document}

\title{\bf Stability properties for some non-autonomous dissipative
phenomena proved by families of Liapunov functionals}
 \author{ {\sc  A. D'Anna   \hspace{30mm} G. Fiore}  \\\\
  Dip. di Matematica e Applicazioni, Fac.  di Ingegneria\\
   Universit\`a di Napoli, V. Claudio 21, 80125 Napoli
       }
 \maketitle

 \vspace{1mm}

{\sc Abstract}: {\small\it We prove some new results regarding the boundedness, stability and attractivity of the solutions
of a class of initial-boundary-value problems characterized
by a quasi-linear third order equation which may contain time-dependent
coefficients. The class includes equations arising in Superconductor Theory,
and in the Theory of Viscoelastic Materials. In the proof we use a family
of Liapunov functionals $W$ depending on two parameters, which we adapt to
the `error', i.e. to the size  $\sigma$ of the chosen neighbourhood of the null solution.
}

 \vspace{1mm}


 \vspace{1mm}

 {\sc A.M.S. Classification}: {\small 35B35 - 35G30}

\bigskip
\noindent
Preprint 08-45 Dip. Matematica e Applicazioni, Universit\`a di Napoli

\sect{Introduction}
In this paper we study the boundedness and stability properties of a large class of initial-boundary-value problems of the form \bea
\label{eq} && \left\{\ba{l} -\varepsilon(t)
u_{xxt}+u_{tt}-C(t)u_{xx}+a'u_t=F(u)-au_t ,\qquad\quad
 x\in]0,\!\pi[,\quad t\!>\!t_0, \\[8pt]
u(0,t)=0, \quad u(\pi,t)=0,  \ea\right.\qquad\qquad  \\[8pt]
&& \quad\:\, u(x,t_0)= u_0(x), \qquad u_t(x,t_0)= u_1(x).\label{eq2}
 \eea
Here $t_0\ge 0$, $\varepsilon\!\in\! C^2(I,I)$,
$C\!\in\! C^1(I,\b{R}^+)$ (with $I\!:=\![0,\infty[$) are functions of $t$,
with $C(t)\!\ge\! \overline{C}\!=\!\mbox{const}\!>\!0$,  the conservative force fulfills $F(0)=0$, so that the equation admits the trivial solution
$u(x,t)\equiv 0$;
$a'=\mbox{const}\!\ge\!0$, $a=a(x,t,u,u_x,u_t,u_{xx})\!\ge\!0$, $\varepsilon(t)\!\ge\! 0$,
so that the corresponding terms are dissipative\footnote{This follows
from the non-positivity of the corresponding terms in the time derivative of the Hamiltonian:
$$
H=\displaystyle\int\limits_0^\pi\!\!dx
\!\left[\frac{\!u_t^2\!+\!C u_x^2}2\!-\!
\int_0^{u(x)}\!\!\!\!\!\!\!F(z)dz\!\right]\qquad\qquad \Rightarrow\qquad
\qquad\dot H=-\int\limits_0^\pi\!\!dx\left[(a\!+\!a')u_t^2\!+\!\varepsilon u_{xt}^2\right]+\int\limits_0^\pi\!\!dx\,\dot C \frac{u_x^2}2.
$$
We also see that the last term is respectively dissipative, forcing if $\dot C$ is negative, positive.
$H$ can play the role of Liapunov functional w.r.t. the reduced norm
$d_{\varepsilon=0}(u,u_t)$.}.

\medskip
Solutions $u$ of such
problems describe a number of physically remarkable
continuous phenomena occurring on a finite space interval.

For instance,
when $F(u)=b \sin u$, $a=0$ we deal with a perturbed
Sine-Gordon  equation which is used 
to describe the classical Josephson effect  \cite{Jos}
in the Theory of Superconductors, which is at the base (see e.g. \cite{LonSco77,BarPat82} of a large number of
advanced developments  both in fundamental research (e.g.
macroscopic effects of quantum physics, quantum computation) and in
applications to electronic devices (see e.g. Chapters 3-6 in
\cite{ChrScoSoe99}): $u(x,t)$ is the phase difference of the macroscopic quantum
wavefunctions describing the Bose-Einstein condensates of
Cooper pairs in two superconductors separated by a very thin and narrow
 dielectric strip (a socalled ``Josephson junction''), 
the dissipative term $(a'\!+\!a)u_t$ is due to Joule effect of the residual 
current across the junction due to single electrons, whereas the third
order dissipative term is due to the surface impedence of the two
superconductors of the strip. Usually the model is considered
with constant (dimensionless)
coefficients $\varepsilon,C,(a'\!+\!a)$, but in fact the latter
depend on other physical parameters like the temperature 
or the voltage difference applied to the junction (see e.g. \cite{LonSco77}),
which can be controlled and varied with time; in a more accurate
description of the model one should take a non-constant 
$a=\beta \cos u$, where $\beta$ also depends on  temperature 
and voltage difference applied and therefore can be varied with time.

Other applications of problem (\ref{eq}-\ref{eq2}) include heat conduction 
at low temperature \cite{MorPayStr90,FlaRio96}, sound propagation in 
viscous gases \cite{Lam32}, 
propagation of plane waves in perfect incompressible and electrically 
conducting fluids \cite{Nar53}, motions of viscoelastic fluids or 
solids \cite{JosRenSau85,Mor56,Ren83}. For instance,
problem (\ref{eq}-\ref{eq2}) with $a=0=a'$
describes \cite{Mor56}
the evolution of the displacement $u(x,t)$ of the section of a rod
from its rest position $x$ in a
Voigt material when an external force $F$ is applied; 
in this case $c^2=E/\rho$, $\varepsilon=1/(\rho\mu)$,
where $\rho$ is the (constant) linear density of the rod at rest,
and $E,\mu$ are respectively
the elastic and viscous constants of the rod, which enter
the stress-strain relation
$\sigma=E\nu+\partial_t \nu/\mu$,
where $\sigma$ is the stress, $\nu$ is the strain. Again, some
of these parameters, like the viscous constant of the rod, may
depend on the temperature of the rod, which can be controlled
and varied with time.

\medskip
The problem (\ref{eq}-\ref{eq2}) considered here generalizes 
those considered in
\cite{DacDan95,DacDan98,DanFio00,DanFio05}, in that
the square velocity $C$ and the dissipative coefficient $\varepsilon$ can
depend on $t$. The physical phenomena just described provide the
motivations for such a generalization.
While  we require $C$ to have a positive lower bound, in order
not to completely destroy the wave propagation effects due to
operator $\partial_t^2-C\partial_x^2$,
we wish to include the cases that $\varepsilon$ goes to
zero as $t\to\infty$, vanishes at some point $t$, or even vanishes identically.
To that end, we consider the $t$-dependent norm
\beq
d^2(\varphi,\psi)\equiv
d_\varepsilon^2(\varphi,\psi)=\int\limits_0^\pi\!\!dx\,[\varepsilon^2(t)
\varphi_{xx}^2\!+\!\varphi_x^2\!+\!\varphi^2\!+\!\psi^2];
\eeq
 $\varepsilon^2$ plays the role of a weight for the
second order derivative term $\varphi_{xx}^2$ so that
for $\varepsilon=0$ this automatically reduces to the proper norm needed
for treating the corresponding second order problem. Imposing the condition
that $\varphi,\psi$ vanish in $0,\pi$ one easily derives that
$|\varphi(x)|,\varepsilon|\varphi_x(x)|\le d(\varphi,\psi) $ for any $x$;
therefore a convergence in the norm $d$ implies also a
uniform (in $x$) pointwise convergence of  $\varphi$ and a
uniform (in $x$) pointwise convergence of  $\varphi_x$ for
$\varepsilon(t)\!\neq\! 0$.
To evaluate the distance of $u$ from the trivial solution
we shall use the $t$-dependent norm
$d(t)\equiv d_{\varepsilon(t)}\big[u(x,t),u_t(x,t)\big]$;
we use the abbreviation $d(t)$ whenever this is not ambiguous.

In section \ref{preliminaries} we state the hypotheses necessary to prove
our results, give
the relevant definitions of boundedness and (asymptotic) stability,
introduce a 2-parameter family
of Liapunov functionals $W$ and tune these parameters in order to prove
bounds for $W,\dot W$. In sections
\ref{stability}, \ref{largestability}
we prove the main results: a theorem of stability and (exponential)
asymptotic stability of the null solution
(section \ref{stability}), under stronger assumptions theorem of eventual
and/or uniform boundedness of the solutions and
eventual and/or exponential asymptotic stability in the large of the null
solution (section \ref{largestability}).
In section \ref{examples} mention some examples to which these results can
be applied.

\sect{Main assumptions, definitions and preliminary estimates}
\label{preliminaries}

For any function $f(t)$ we denote $\overline{f}=\inf_{t\!>\!0}f(t)$,
$\overline{\overline{f}}=\sup_{t\!>\!0}f(t)$.
We assume that there exist constants $A\!\ge\!0$, $\tau\!>\!0$,
$k\!\ge\!0$, $\rho\!>\!0$, $\mu\!>\!0$ such that
\bea
&& F(0)\!=\!0\quad\:\: \&\quad\:\: F_z(z)\!\le\! k\quad\quad\:\:
\mbox{ if } |z|\!<\!\rho.
\label{condi3}\\[8pt]
&& \overline{C}\!\ge\!k,\qquad\quad
C\!-\!\dot\varepsilon\!\ge\!\mu(1\!+\!\varepsilon),\qquad\quad
\mu\!+\!\frac{\overline{C}}2
\!-\!2k\!>\!0,  \qquad\quad \overline{\ddot\varepsilon}\!>\!-\infty.
\label{condi2}\\[8pt]
&& 0\le a\!\le\! A d^\tau(u,u_t),\qquad\qquad a'\!+\!
\frac{\overline{\varepsilon}}2\!>\!0\label{condi1}
\eea
We are not excluding the following cases: $\varepsilon(t)=0$ for some $t$,
$\varepsilon\stackrel{t\!\to\!\infty}{\longrightarrow}0$,
$\varepsilon(t)\equiv 0$,
$\varepsilon\stackrel{t\!\to\!\infty}{\longrightarrow}\infty$
[in view of (\ref{condi2})$_2$ the latter condition requires also
$C\stackrel{t\!\to\!\infty}{\longrightarrow}\infty$];
but by condition (\ref{condi1})$_2$ at least one of the dissipative terms
must be nonzero. Eq. (\ref{condi3}) implies
\beq
\int_0^{\varphi}\!\!\!F(z)dz\!\le\! k\frac {\varphi^2}2,\qquad\qquad
\varphi F(\varphi)\!\le\!
k\varphi^2\quad\qquad \mbox{if } |\varphi|\!<\!\rho.   \label{conseq}
\eeq
We shall consider also the cases that, in addition to (\ref{condi3}),
either one of the following inequalities [which are stronger than
(\ref{conseq})] holds:
$$
\int_0^{\varphi}\!\!\!F(z)dz\!\le\! 0,\qquad\qquad
\varphi F(\varphi)\!\le\! 0\quad\qquad \mbox{if } |\varphi|\!<\!\rho.   \eqno{(\ref{conseq}')}
$$

\bigskip
To formulate our results we need the following definitions.
Fix once and for all $\kappa\in\b{R}$, $\xi\!>\!0$ and let $I_\kappa:=[\kappa,\infty[$, $d(t):= d_{\varepsilon(t)}\big[u(x,t),u_t(x,t)\big]$.

\medskip
{\sc Definition 2.1}. The solution $u(x,t)\equiv 0$ of (\ref{eq}) is
stable  if for any $\sigma\!\in\!]0,\xi]$ and  $t_0\!\in\!I_\kappa$ there exits a
$\delta(\sigma,t_0)> 0$ such that
$$
d(t_0)<\delta(\sigma,t_0)\qquad
\qquad\Rightarrow\qquad \qquad d(t)<\sigma\:\quad\forall t\ge t_0.
$$
If $\delta$ can be chosen independent of $t_0$,
$\delta=\delta(\sigma)$, $u(x,t)\equiv 0$  is uniformly
stable.

\bigskip

 {\sc Definition 2.2}. The solution $u(x,t)\equiv 0$ of (\ref{eq}) is  asymptotically stable if it is  stable and moreover for any
$t_0\!\in\!I_\kappa$ there exists a $\delta(t_0)\!>\!0$  such that
$d(t_0)\!<\!\delta(t_0)$ implies $d(t)\to 0$ as $t\to\infty$, namely
for any $\nu\!>\!0$ there exists a $T(\nu,t_0,u_0,u_1)> 0$
such that
$$
d(t_0)<\delta(t_0)\qquad \qquad\Rightarrow\qquad
\qquad d(t)<\nu\:\quad\forall t\ge t_0+T.
$$
The solution $u(x,t)\equiv 0$ is uniformly asymptotically stable if it is uniformly stable and moreover $\delta,T$ can be chosen independent of $t_0, u_0,u_1$, i.e. $d(t)\to 0$
as $t\to\infty$ uniformly in $t_0,u_0,u_1$.

\bigskip

{\sc Definition 2.3}. The solutions of (\ref{eq}) are  eventually
uniformly bounded  if for any $\delta>0$ there exist a
$s(\delta)\geq 0$ and a $\beta(\delta)>0$ such that if  $t_0\geq
s(\delta)$, $d(t_0)\leq \delta$, then
$d(t)<\beta(\delta)$ for all $t\geq t_0$. If $s(\delta)=0$
the solutions of (\ref{eq}) are uniformly bounded.

\bigskip

{\sc Definition 2.4}. The solutions of (\ref{eq}) are
 bounded  if for any $\delta>0$ there exist a
 $\tilde\beta(\delta,t_0)>0$ such that if  $d(t_0)\leq \delta$, then
$d(t)<\tilde\beta(\delta,t_0)$ for all $t\geq t_0$.

\bigskip

 {\sc Definition 2.5}. The solution $u(x,t)\equiv 0$ of (\ref{eq})
 is eventually exponential-asymptotically stable in the large if
for any $\delta>0$ there are a nonnegative constant
$s(\delta)$ and  positive constants $D(\delta),
E(\delta)$ such that if $t_0\ge s(\delta)$,  $d(t_0)\leq \delta$, then \beq
   d(t)\leq D(\delta)
   \exp \left[-E(\delta)(t-t_0) \right] d(t_0), \qquad   \forall
   t\geq t_0.                                     
\eeq
If $s(\delta)=0$ then $u(x,t)\equiv 0$
 is  exponential-asymptotically stable in the large.

\bigskip

 {\sc Definition 2.6}. The solution $u(x,t)\equiv 0$ of (\ref{eq})
 is (uniformly) exponential-asymptotically stable if there exist positive constant $\delta, D,E$ such that \beq
 d(t_0)<\delta\qquad \qquad\Rightarrow\qquad
\qquad  d(t)\leq D
   \exp \left[-E(t-t_0) \right] d(t_0), \qquad   \forall
   t\geq t_0.                                     
\eeq

\bigskip

 {\sc Definition 2.7}.  The solution $u(x,t)\equiv 0$ of (\ref{eq}) is  asymptotically stable in the large if it is  stable and moreover for any
$t_0\!\in\!I_\kappa$,  $\nu,\alpha>0$ there exists
$T(\alpha,\nu,t_0,u_0,u_1)> 0$
such that
$$
d(t_0)<\alpha\qquad \qquad\Rightarrow\qquad
\qquad d(t)<\nu\:\quad\forall t\ge t_0+T.
$$

\bigskip

We recall Poincar\'e inequality, which easily follows from Fourier analysis:
\beq
\phi\in C^1(]0,\pi[),\quad\phi(0)=0,
\quad\phi(\pi)=0,\qquad\Rightarrow \qquad \int^\pi_0 \!\!\!
dx\,\phi_x^2(x)\ge \int^\pi_0 \!\!\! dx\,\phi^2(x). \label{poinc}
\eeq

We introduce the non-autonomous family of Liapunov functionals
\bea
\label{321}
W\equiv W(\varphi,\psi,t;\gamma,\theta)&:=&\int_0^\pi\!
\frac{1}{2}\!\Big\{\!\gamma\psi^2\!+\!(\varepsilon
\varphi_{xx}\!-\!\psi)^2\!\!+
[C(1\!+\!\gamma\!)\!-\!\dot\varepsilon\!+\!\varepsilon(a'\!+\!\theta)]\varphi_x^2
\\ &&\qquad\quad
+a'\theta\varphi^2\!+\!2\theta\varphi\psi \!-\!
2(1\!+\!\gamma)\!\!
\int_0^{\varphi(x)}\!\!\!\!\!\!\!F(z)dz\!\Big\}dx\nonumber
\eea
where $\theta,\gamma$ are for the moment  unspecified positive
parameters. $W$ coincides with the Liapunov functional of
\cite{DacDan95} for constant $\varepsilon,C$ and $\gamma=3$,
$\theta=a'$. Let $W(t;\gamma,\theta)\!:=\!W(u,u_t,t;\gamma,\theta)$.
Using (\ref{eq}), from (\ref{321})  one finds
\bea
\dot W(t;\gamma,\theta)&=&\int_0^\pi\!
\!\left\{\!(\varepsilon u_{xx}\!-\!u_t)(\varepsilon u_{xxt}
\!-\!u_{tt}\!+\!\dot\varepsilon u_{xx}) \!+\![\dot C(1\!+\!\gamma\!)\!-\!\ddot\varepsilon\!+\!\dot \varepsilon(a'\!+\!\theta)]
\frac{u_x^2}2\right.\nn &&\qquad \left.\!+\!
[C(1\!+\!\gamma\!)\!-\!\dot\varepsilon\!+\!\varepsilon(a'\!+\!\theta)]u_xu_{xt}
\!+\!a'\theta uu_t\!+\!\theta u_t^2\!+\!(\gamma u_t\!+\!\theta u)u_{tt} \!-\!
(1\!+\!\gamma)F(u) u_t\right\}dx
\nn &=&\int_0^\pi\!
\!\left\{\!(\varepsilon u_{xx}\!-\!u_t)[(a\!+\!a')u_t \!-\! Cu_{xx}\!-\!F(u) \!+\!\dot\varepsilon u_{xx}] \!+\![\dot C(1\!+\!\gamma\!)\!-\!\ddot\varepsilon\!+\!\dot \varepsilon(a'\!+\!\theta)]
\frac{u_x^2}2\!-\!
[C(1\!+\!\gamma\!)\!-\!\dot\varepsilon\right.\nn &&\left.\!+\!\varepsilon(a'\!+\!\theta)]u_{xx}u_t
\!+\!a'\theta uu_t\!+\!\theta u_t^2\!+\!(\gamma u_t\!+\!\theta u)[Cu_{xx}\!+\!\varepsilon u_{xxt}\!+\!F(u)\!-\!(a\!+\!a')u_t]  \!-\!(1\!+\!\gamma)F(u) u_t\right\}dx \nn &= &
\int_0^\pi\!
\!\left\{\!\varepsilon u_{xx}[(\dot\varepsilon\!-\! C)\!-\!F(u)]u_{xx}\!+\!u_t[\varepsilon u_{xx}(a\!+\!a')\!-\!(a\!+\!a')u_t \!+\! Cu_{xx}\!+\!F(u) \!-\!\dot\varepsilon u_{xx}\!-\!
C(1\!+\!\gamma\!)u_{xx}\right.\nn &&\qquad\!+\!\dot\varepsilon u_{xx}\!-\!\varepsilon(a'\!+\!\theta)u_{xx}\!+\!a'\theta u\!+\!\theta u_t\!+\!
\gamma Cu_{xx}\!+\!\gamma\varepsilon u_{xxt}\!+\!\gamma F(u)\!-\!(a\!+\!a')\gamma u_t\!-\!\theta(a\!+\!a')u\nn &&\qquad \left.
\!-\!(1\!+\!\gamma)F(u)]\!+\!\theta u[Cu_{xx}\!+\!\varepsilon u_{xxt}\!+\!F(u)]
\!+\![\dot C(1\!+\!\gamma\!)\!-\!\ddot\varepsilon\!+\!\dot \varepsilon(a'\!+\!\theta)]\frac{u_x^2}2\right\}dx
\nn&= &
\int_0^\pi\!
\!\Big\{\!\varepsilon [(\dot\varepsilon\!-\! C)u_{xx}\!-\!F(u)]u_{xx}\!+\!u_t[\varepsilon a u_{xx}  \!-\!(a\!+\!a')(1\!+\!\gamma\!)u_t
\!-\!\varepsilon\theta u_{xx}
\nn &&\qquad \left.\!+\!\theta u_t
\!+\!\gamma\varepsilon u_{xxt}
\!-\!a\theta u]\!+\!\theta u[Cu_{xx}\!+\!\varepsilon u_{xxt}\!+\!F(u)]
\!+\![\dot C(1\!+\!\gamma\!)\!-\!\ddot\varepsilon\!+\!\dot \varepsilon(a'\!+\!\theta)]
\frac{u_x^2}2\right\}dx
\nn&= & -\int_0^\pi \!\!\left\{\varepsilon(C\!-\!\dot\varepsilon) u^2_{xx}\!+\!
\left[(a\!+\!a')(1\!+\!\gamma)\!-\!\theta\right]u_t^2\!+\!
\left[2\theta
C\!+\!\ddot\varepsilon\!-\!\dot\varepsilon(a'\!+\!\theta)\!-\!(1\!+\!\gamma)\dot
C\right]\frac{u_x^2}2 \!+\!   \varepsilon\gamma
u^2_{xt}\right.\nn[8pt] &&\qquad \qquad\qquad\left. \!+\!\theta
auu_t \!-\!\theta u F(u)\!+\!\varepsilon[-\!a u_t\!+\!F(u)]u_{xx}
\right\}dx
\eea

\subsection{Upper bound for $\dot W$}

After some rearrangement of terms and
integration by parts of the last term, we obtain
\bea &&\dot W=-\int_0^\pi \!\!\left\{\varepsilon\gamma
u^2_{xt}\!+\!
\left[(a\!+\!a')(1\!+\!\gamma)\!-\!\theta\!-\!\varepsilon\frac{a^2}{C\!-\!\dot\varepsilon}\!-\!\theta\frac{a^2}C
\right]u_t^2\!+\!\varepsilon(C\!-\!\dot\varepsilon)\left[\frac{a}{C\!-\!\dot\varepsilon}u_t\!-\!\frac{u_{xx}}2\right]^2
\!+\!\frac 34\varepsilon(C\!-\!\dot\varepsilon) u^2_{xx}
\right.\nn[8pt] &&\left. \!+\!\left[
C\left(\frac {\theta}2\!-\!a'\right)\!+\!\ddot\varepsilon\!+\!(C\!-\!
\dot\varepsilon)(a'\!+\!\theta)
\!-\!(1\!+\!\gamma)\dot C\!-\!2\varepsilon
F_u\right]\frac{u_x^2}2\!+\!\frac{\theta C}4(u_x^2\!-\!u^2)
\!+\!\frac{\theta C}4\left[u\!+\!\frac {2a} C u_t\right]^2
\!-\!\theta uF(u)\right\}dx
\nonumber \eea

Using (\ref{poinc}) with $\phi(x)=u_t(x,t),u(x,t)$ we thus find, provided
$|u|\!<\!\rho$, $\theta\!>\!2a'$, $\mu(a'\!+\!\theta)\!>\!2 k$
 \bea
&&\dot W \!\le\! -\!\int_0^\pi\!
\!\!\left\{\!\left[\varepsilon\gamma\!+\!(a\!+\!a')(1\!+\!\gamma)\!-\!\theta\!-\!a^2\!\left(\!\frac 1{\mu}
\!+\!\frac{\theta}C\!\right)\!\right]\!u_t^2\!+\!\frac 34 \mu\varepsilon^2
u^2_{xx}\!+\!\right.\nn &&
\qquad \qquad\:\left.\left[C\!\left(\!\frac {\theta}2\!-\!a'\!\right)
\!\!+\!\ddot\varepsilon\!+\!\mu(1\!+\!\varepsilon)(a'\!+\!\theta)\!-\!(1\!+\!\gamma)\dot
C\!-\!2\varepsilon k\right]\!\frac{u_x^2}2  \!-\!\theta ku^2\!
\right\}\!dx\nn &&\!\!\le\!\! -\!\int_0^\pi\!
\!\!\left\{\!\left[\overline{\varepsilon}\gamma\!+\!(a\!+\!a')(1\!+\!\gamma)
\!-\!\theta\!-\!a^2\!\left(\!\frac 1{\mu}
\!+\!\frac{\theta}{\overline{C}}\!\right)\right]\!u_t^2
\!+\!\frac 34 \mu\varepsilon^2 u^2_{xx}\!+\!\right.\nn &&
\qquad \qquad\:\left.\left[\overline{C}\!\left(\!\frac {\theta}2\!-\!a'\!\right)\!\!+\!\overline{{\ddot\varepsilon}}\!+\!\mu(a'\!+\!\theta)
\!+\![\mu(a'\!+\!\theta)\!-\!2 k]\varepsilon
\!-\!(1\!+\!\gamma)\dot C\!-\!2 k\theta\right]\!\frac{u_x^2}2\!
\right\}\!dx.\qquad \label{ineq0}
\eea

To fix $\theta$ we now assume that there exists $\bar t(\gamma)\!\in\! [0,\infty[$
such that
\beq
\dot C(1+\gamma)\!<\!1\quad \mbox{for }t\!>\!\bar t,\qquad \qquad
\dot C(1+\gamma)\!\ge\!1\quad \mbox{for }0\!\le\!t\!\le\!\bar t.\label{Cdotip}
\eeq
This is clearly satisfied
with $\bar t(\gamma)\!\equiv\! 0$ if $\dot C\le 0$, whereas it is satisfied
with some $\bar t(\gamma)\!\ge\!0$ if $\dot C\stackrel{t\!\to\!\infty}{\longrightarrow}0$.
Correspondingly, we choose
\beq
\theta>\theta_1:=\max\left\{2a',\frac{2k}
{\mu}\!-\!a' ,\frac{5\!-\!\overline{\ddot
\varepsilon}\!-\! a'(\mu\!-\!\overline{C})}
{\mu\!+\!\overline{C}/2\!-\!2k}\right\}   \label{thetadef}
\eeq
Then for all $t> \bar t$
\beq \theta\!\left(\!\mu\!+\!\frac {\overline{C}}2\!-\!2k\!\right)\!\!+\![\mu(a'\!+\!\theta)\!-\!2 k]\overline{\varepsilon}\!+\!\overline{\ddot\varepsilon}\!-\!(1\!+\!\gamma)
\dot C\!+\!a' (\mu\!-\! \overline{C})> 4.\label{ineq1}
\eeq
Next, provided $d(u,u_t)\!\le\!  \sigma\!<\! \rho$ we choose
\beq
\gamma>\gamma_1(\sigma):=\frac {1\!+\!\theta}{a'\!+\!\overline{\varepsilon}}
+\gamma_{32}\sigma^{2\tau}\qquad\qquad \gamma_{32}:=\frac {A^2}{(a'\!+\!\overline{\varepsilon})}
\left(\frac 1{\mu}\!+\!\frac{\theta}{\overline{C}}\right),
\label{defgamma1}
\eeq
what implies, for $d\le \sigma$,
\bea
&&\overline{\varepsilon}\gamma\!+\!(a\!+\!a')(1\!+\!\gamma)\!-\!\theta
\!-\!a^2\!\left(\!\frac 1{\mu}\!+\!\frac{\theta}{\overline{C}}\!\right)=a\!+\!a'\!+(a\!+\!a'\!+\!\overline{\varepsilon})\gamma\!-\!\theta\!-\!a^2\!\left(\!\frac 1{\mu}\!+\!\frac{\theta}{\overline{C}}\!\right)\nn
&&\ge
a'\!+\!\frac {a\!+\!a'\!+\!\overline{\varepsilon}}{a'\!+\!\overline{\varepsilon}}\left[(1\!+\!\theta)
+A^2\left(\frac 1{\mu}\!+\!\frac{\theta}{\overline{C}}
\right)\sigma^{2\tau}\right]\!-\!\theta\!-
\!A^2\!\left(\!\frac 1{\mu}\!+\!\frac{\theta}{\overline{C}}\!\right)d^{2\tau}
\ge 1\!+\!a'.          \label{ineq2}
\eea
Equations (\ref{ineq0}), (\ref{ineq1}) and (\ref{ineq2}) imply
for all $t\ge \bar t$
\bea
\dot W(u,u_t,t;\gamma,\theta) &\!\!\le\!\!& -\!\int_0^\pi\!
\!\!\left\{\!\left[\overline{\varepsilon}\gamma\!+\!(a\!+\!a')(1\!+\!\gamma)\!-\!\theta\!-\!a^2\!\left(\!\frac 1{\mu}
\!+\!\frac{\theta}{\overline{C}}\!\right)\right]\!u_t^2
\!+\!\frac 34 \mu\varepsilon^2 u^2_{xx}\!+\!\right.\nn &&
\qquad\:\left.
\left[\theta\!\left(\!\mu\!+\!\frac {\overline{C}}2\!-\!2k\!\right)\!\!+\![\mu(a'\!+\!\theta)\!-\!2 k]\overline{\varepsilon}\!+\!
\overline{\ddot\varepsilon}\!-\!(1\!+\!\gamma)\dot C\!+\!a'
(\mu\!-\! \overline{C})\right]\!\frac{u_x^2+u^2}4\!
\right\}\!dx\nn[8pt] &\!\!<\!\!& -\eta\, d^2(t),\qquad\qquad\qquad
\eta:=\min\left\{1,3 \mu/4\right\}                  \label{Ineq1}
\eea
provided $0\!<\!d(t)\!<\! \sigma$.
If, in addition to (\ref{condi1}) with $k>0$, the inequality
(\ref{conseq}') [which is stronger than (\ref{conseq})] holds,
then it is easy to check that we can avoid assuming (\ref{condi2})$_3$ and
obtain again the previous inequality replacing $k\to 0$
in the definition (\ref{thetadef}) of $\theta_1$.

\medskip
{\bf Remark 1.} One can check that if we had adopted the same
Liapunov functional as in \cite{DanFio00,DanFio05} formulae (4.2),
i.e. $W$ of (\ref{321}) with $\theta\!=\!0\!=\!a'$, we would have
not been able to obtain (\ref{Ineq1}) (which is essential to prove
the asymptotic stability of the null solution) in a number of
situations, e.g. if $\varepsilon\!\to\!0$ sufficiently fast as
$t\!\to\! \infty$.

\subsection{Lower bound for $W$}

From the definition (\ref{321}) it immediately follows
\bea
&& W(\varphi,\psi,t;\gamma,\theta) =\int\limits_0^\pi\!
\frac{1}{2}\!\left\{\left(\gamma\!-\!\theta^2\!-\!\frac
12\right)\psi^2\!+\!\frac{(\varepsilon \varphi_{xx}\!-2\!\psi)^2}4
\!+\!\frac{(\varepsilon \varphi_{xx}\!-\!\psi)^2}2\!+\!
\varepsilon^2\frac{\varphi_{xx}^2}4\right.\nn[8pt] && \left.\qquad
\qquad+\![C(1\!+\!\gamma)\!-\!\dot\varepsilon\!+\!\varepsilon
(a'\!+\!\theta)]\varphi_x^2 \!+\!(a'\theta\!-\!1)\varphi^2\!+\!\left[\theta\psi\!+\!\varphi\right]^2\!  \!-\! 2(1\!+\!\gamma)\!\!
\int_0^{\varphi(x)}\!\!\!\!\!\!F(z)dz\right\}dx     \label{Wlower}
\eea

Using (\ref{condi2})$_2$, (\ref{conseq}) and (\ref{poinc}) with
$\phi(x)=\varphi(x)$ we find for $|\varphi|\!<\!\rho$ \bea &&
W\!\ge\! \int\limits_0^\pi\!
\frac{1}{2}\!\left\{\left(\gamma\!-\!\theta^2\!-\!\frac 12\right)
\psi^2\!+\!\varepsilon^2\frac{\varphi_{xx}^2}4\!+\!\left[
(C\!-\!k)\gamma\!+\mu\!\!+\!(\mu\!+\!a'\!+\!\theta)\varepsilon\right]
\varphi_x^2\!+\!\left[a'\theta\!-\!1\!-\! k\right]
\varphi^2\right\}dx \nn && \!\ge\! \int\limits_0^\pi\!
\frac{1}{2}\!\left\{\left(\gamma\!-\!\theta^2\!-\!\frac 12\right)
\psi^2\!+\!\varepsilon^2\frac{\varphi_{xx}^2}4\!+\!\left[(
\overline{C}\!-\!k)\gamma\!+\!\mu\!\!+\!\left(\mu\!+\!a'\!+\!\frac{\theta}{2}\right)\overline{\varepsilon}\right]
\varphi_x^2\!+\!\left[\left(a'\!+\!\frac{\overline{\varepsilon}}{2}\right)\theta
\!-\!1\!-\!k\right] \varphi^2\right\}dx.\qquad\quad  \eea
Choosing
\beq
\theta>\theta_2:=\max\left\{\theta_1,
\frac{k\!+\!5/4}{a'\!+\!\overline{\varepsilon}/2}\right\},\qquad\qquad\gamma\ge\gamma_2(\sigma):=\gamma_1(\sigma)\!+\!\theta^2\!+\!1  \label{defthetagamma2}
\eeq
we find that for $d\le \sigma$
\beq
W(\varphi,\psi,t;\gamma,\theta) \ge \chi \:
d^2(\varphi,\psi),\qquad\qquad \chi:=\frac 12\min\left\{\frac
12,(\overline{C}\!-\!k)\gamma\!+\!\mu\!\!+\!\left(\mu\!+\!a'\!+\!\frac{\theta}{2}\right)\overline{\varepsilon}   \right\}>0. \label{Ineq2}
\eeq
(Note that $\chi\le 1/4$). If, in
addition to (\ref{condi3}) (with some $k\!>\!0$), the inequality
(\ref{conseq}')$_1$ holds, then it is easy to check that we obtain
(\ref{Ineq2}) [with the replacement $k\to 0$ in the definition of
$\chi$] by choosing $\theta,\gamma$ as in (\ref{defthetagamma2}),
but replacing $k\to 0$ there.

Finally, we note that if $\tau\!=\!0$ in (\ref{condi1}), i.e. $a\!\le\! A=$const, then $\gamma, \bar t(\gamma)$ are independent of $\sigma$.

\subsection{Upper bound for $W$}

As argued in \cite{DacDan95},
\[
\left|\int_0^\varphi F(z)dz \right|=
\left|\int_0^\varphi dz\int_0^\zeta F_\zeta(\zeta)d\zeta \right|=
  \left|\int_0^\varphi F_\zeta(\zeta)(\varphi-\zeta)d\zeta \right|.
\]
 Consequently, introducing the non-decreasing funtion
\[
m(r):= \max \left\{|F_\zeta(\zeta)| \ : \  |\zeta |\le r\right\}
\]
and in view of the inequality $|\varphi|\le d(\varphi,\psi)$ we obtain
\beq                                                 \label{327}
   \left|\int_0^\varphi F(z)dz \right| \le m(|\varphi |) \frac{\varphi^2}2
\leq m(d) \frac{d^2}2. \eeq Thus, from definition (\ref{321}) and
the inequalities
$-2\epsilon\varphi_{xx}\psi\!\le\!\epsilon^2\varphi^2_{xx}\!+\!\psi^2$,
$2\theta\varphi\psi\!\le\!\theta(\varphi^2\!+\!\psi^2)$,
(\ref{condi2})$_3$
 we easily find
\bea
&& W(\varphi,\psi,t;\gamma,\theta)\!\le\!  \int\limits_0^\pi\!
\frac{1}{2}\!\left\{\!(\gamma\!+\!2\!+\!\theta)\psi^2\!+\!2\varepsilon^2
\varphi_{xx}^2\!\!+\!\left[C(1\!+\!\gamma\!)\!-\!\dot\varepsilon\!+\!\varepsilon(a'\!+\!\theta)\right]\varphi_x^2\!+\!(a'\!+\!1)\theta\varphi^2\right\}dx
\!+\! (1\!+\!\gamma)m(d)\frac{d^2}2\qquad\nn && \qquad
\!\le\!\int\limits_0^\pi\!
\frac{1}{2}\!\left\{\!(\gamma\!+\!2\!+\!\theta)\psi^2\!+\!2\varepsilon^2
\varphi_{xx}^2\!\!+\!\left[C\gamma\!+\!(C
\!-\!\dot\varepsilon)\!\left(1\!+\!\frac{a'\!+\!\theta}{\mu}\right)\!\right]\varphi_x^2\!+\!(a'\!+\!1)\theta\varphi^2\right\}dx
\!+\! (1\!+\!\gamma)m(d)\frac{d^2}2 .\nonumber
\eea

Choosing
\beq\ba{l}
\gamma\ge\gamma_3(\sigma):=\gamma_2(\sigma)\!+\!1\!+\!
\frac{a'\!+\!\theta}{\mu}\!+\!(a'\!+\!1)\theta
=\gamma_{31}+\gamma_{32}\sigma^{2\tau},\\[12pt]
\mbox{where }\quad\gamma_{31}:=\frac {1\!+\!\theta}{a'\!+\!\overline{\varepsilon}}
\!+\!\theta^2\!+\!2\!+\!
\frac{a'\!+\!\theta}{\mu}\!+\!(a'\!+\!1)\theta \ea\label{defthetagamma3}
\eeq
and setting
\beq
g(t)\!:=\!C(t)\!-\!\dot\varepsilon(t)/2\!+\! 1\!>\!1, \qquad\qquad
B^2(d):=\left[1\!+\! m(d)\right]d^2  \label{defgB}
\eeq
we find that for $d\le \sigma$
\bea
W(\varphi,\psi,t;\gamma,\theta)&\!\le\! & \int\limits_0^\pi\!
\frac{1}{2}\!\left[\!(\gamma\!+\!2\!+\!\theta)\psi^2\!+\!2\varepsilon^2
\varphi_{xx}^2\!\!+\!\gamma\left(2C
\!-\!\dot\varepsilon\right)\varphi_x^2\!+\!\gamma\varphi^2\right]dx
\!+\! (1\!+\!\gamma)m(d)\frac{d^2}2  \nn &\!\le\! & \left[2\gamma
g(t) \!+\! (1\!+\!\gamma)m(d)\right]\frac{d^2}2 \!\le\!
\left(1\!+\!\gamma\right)\left[g(t) \!+\! m(d)\right]d^2  \nn[8pt]
&\!\le\! & \left[1\!+\!\gamma(\sigma)\right]g(t) B^2(d).\label{Ineq3}
\eea
The map $d\!\in\![0,\infty[\to
B(d)\!\in\![0,\infty[$ is  continuous and increasing, therefore also invertible. Moreover, $B(d)\ge d$.


\sect{Asymptotic stability of the null solution}
\label{stability}

\begin{theorem}
Assume that conditions (\ref{condi1}-\ref{condi3}) are fulfilled.
Then the null solution $u(x,t)$ of (\ref{eq}) is stable
if one of the following conditions is fulfilled:
\bea
&&\dot C\le 0,\qquad\qquad\forall t\in I,\label{CdotCondi1}\\[8pt]
&&\dot C\stackrel{t\!\to\!\infty}{\longrightarrow}0; \label{CdotCondi2}
\eea
the  stability is  uniform if the function $g(t)$ defined by
(\ref{defgB}) fulfills  $\overline{\overline g}<\infty$.
The $\xi$ appearing in Def. 2.1 is a suitable positive constant,
more precisely $\xi\in]0,\rho]$ if $\rho<\infty$.
The null solution is asymptotically stable if, in addition,
\beq
\int^\infty_0 \frac{dt}{g(t)}=\infty,          \label{AsymStabCondi}
\eeq
and uniformly exponential-asymptotically stable if
$\overline{\overline g}<\infty$.
\label{thm1}
\end{theorem}

\bp~
As a first step, we analyze the behaviour of
$$
\frac{\sigma^2}{1\!+\!\gamma_3(\sigma)}
=\frac{\sigma^2}{1\!+\!\gamma_{31}\!+\!\gamma_{32}\sigma^{2\tau}}=:r^2(\sigma).
$$
The positive constants $\gamma_{31},\gamma_{32}$, defined in
(\ref{defthetagamma3}), are independent of
$\sigma, t_0$. The function $r(\sigma)$ is an increasing and
therefore invertible map  $r\!:\![0,\sigma_M[\to [0,r_M[$, where:
\beq
\ba{lll}
\sigma_M\!=\!\infty,\qquad \qquad & r_M\!=\!\infty,\qquad\qquad & \mbox{if }\:\tau\!\in\![0,1[,\\[8pt]
\sigma_M\!=\!\infty\qquad \qquad & r_M\!=\!1/\sqrt{\gamma_{32}},\qquad\qquad & \mbox{if }\:\tau\!=\!1,\\[8pt]
\sigma_M^{2\tau}:=(1\!+\!\gamma_{31})/\gamma_{32}(\tau\!-\!1),
\qquad \qquad & r_M\!=\![(\tau\!-\!1)/(1\!+\!\gamma_{31})]^{\frac{\tau\!-\!1}{2\tau}}/\sqrt{\tau}\gamma_{32}^{\frac 1{2\tau}},\qquad\qquad & \mbox{if }\:
\tau\!>\!1,\ea                         \label{defsigmaM}
\eeq
[in the latter case $r(\sigma)$ is decreasing beyond $\sigma_M$].

Next, let $\xi\!:=\!\min\{\sigma_M,\rho\}$ if the rhs is finite,
otherwise choose $\xi\in\b{R}^+$; we shall consider an
``error'' $\sigma\!\in]0,\xi[$. We define
\beq
\delta(\sigma,t_0):=B^{-1}\!\left[r(\sigma)\frac{\sqrt{\chi}}{\sqrt{g(
t_0)}}\right], \qquad \qquad \kappa:=\bar t[\gamma_3(\xi)]. \label{defdelta}
\eeq
$\delta(\sigma,t_0)$
belongs to $]0,\sigma[$, because $B(d)\ge d$ implies
$B^{-1}\!\left[r(\sigma)\sqrt{\chi}/\sqrt{g(t_0)}\right]\le
\sqrt{\chi} \sigma\le \sigma/2$ and is an increasing function of $\sigma$.
The function $\bar t(\gamma)$ was defined in
(\ref{Cdotip});  $\bar t[\gamma_3(\sigma)]\!\le\!\kappa$
as the function $\bar t[\gamma_3(\sigma)]$
is non-decreasing. Mimicing an argument
of \cite{DanFio05}, we can show that for any $t_0\ge \kappa$
\beq
d(t_0)<\delta(\sigma,t_0)\qquad \qquad \Rightarrow \qquad \qquad d(t) <\sigma
\qquad\forall t\ge t_0.  \label{subtesi}
\eeq
{\it Ad absurdum}, assume that there
exists a finite $t_1\!>\!t_0$ such that (\ref{subtesi})  is fulfilled
for all $t\in[t_0,t_1[$, whereas
\beq
d(t_1)=\sigma.                           \label{absurdum}
\eeq
The negativity of the rhs(\ref{Ineq1}) implies
that $W(t)\equiv W[u,u_t,t;\gamma_3(\sigma),\theta]$ is a decreasing
function of $t$ in $[t_0,t_1]$. Using (\ref{Ineq2}), (\ref{Ineq3})
we find the following contradiction with (\ref{absurdum}): \bea
&& \chi d^2\!(t_1)\le W(t_1)< W(t_0)\le
\left[1\!+\!\gamma_3(\sigma)\right]g(t_0)
B^2\left[d(t_0)\right]<\left[1\!+\!\gamma_3(\sigma)\right]g(t_0)
B^2\!\left(\delta\right)\nn &&
=\left[1\!+\!\gamma_3(\sigma)\right]g(t_0)
\left\{B\left[B^{-1}\left(\sigma\frac{\sqrt{\chi}}{\sqrt{[1\!+\!\gamma_3(\sigma)]g(t_0)}}\right)\right]\right\}^2=\chi\sigma^2.\nonumber
\eea

Eq. (\ref{subtesi}) amounts to the stability of the
null solution; if
$\overline{\overline g}<\infty$ we obtain the uniform stability
replacing (\ref{defdelta})$_1$
by $\delta(\sigma)\!:=\!B^{-1}\!\left[r(\sigma)\sqrt{\chi}/\sqrt{\overline{\overline g}}\right]$.

Let now $\delta(t_0):=\delta(\xi,t_0)$. By (\ref{subtesi}) and the monotonicity
of $\delta(\cdot,t_0)$ we find that for any $t_0\ge \kappa$
\beq
d(t_0)<\delta(t_0)\qquad \qquad \Rightarrow \qquad \qquad d(t) <\xi
\qquad\forall t\ge t_0.  
\eeq
Choosing $W(t)\equiv W[u,u_t,t;\gamma_3(\xi),\theta]$,
(\ref{Ineq3}) becomes
\beq W(t)\le h(\xi)g(t)d^2(t),  \qquad \qquad \qquad h(\xi):=
\left[1\!+\!\gamma_3(\xi)\right]\left[1 \!+\!
m(\xi)\right], \label{Ineq3'}
\eeq
which
together with (\ref{Ineq1}), implies $\dot W(t)\le -\eta W(t)/[h
g(t)]$ and (by means of the comparison principle \cite{Yos66})
$W(t)<W(t_0)\exp\left[-\eta \int^t_{t_0}dz/[hg(z)]\right]$, whence
$$
d^2(t)\le \frac{W(t)}{\chi}<
 \frac{W(t_0)}{\chi} \exp\!\left[-\frac \eta {h}\!\int\limits^t_{t_0}\!\!\frac{dz}{g(z)}\right]\le
\frac{hg(t_0)}{\chi}d^2(t_0) \exp\!\left[-\frac \eta {h}\!\int\limits^t_{t_0}\!\!\frac{dz}{g(z)}\right]
<\frac{h(\xi)g(t_0)}{\chi}\xi^2 \exp\!\left[-\frac \eta {h(\xi)}\!\int\limits^t_{t_0}\!\!\frac{dz}{g(z)}\right]
$$
Condition (\ref{AsymStabCondi}) implies that the exponential goes to zero
as $t\to\infty$, proving the asymptotic stability of the null solution;
if $\overline{\overline g}\!<\!\infty$ we can replace $g(t_0), g(z)$
by $\overline{\overline g}$ in the last but one inequality and obtain
$$
d^2(t)<\frac{h(\xi)\overline{\overline g}}{\chi} \exp\left[-\frac {\eta} {h(\xi)\overline{\overline g}}(t\!-\!t_0) \right]d^2(t_0),
$$
which proves the uniform exponential-asymptotic stability of the null solution (just set $\delta\!=\!B^{-1}\!\left[r(\xi)\sqrt{\chi}/\sqrt{\overline{\overline g}}\right]$, $D\!=\!\sqrt{h(\xi)\overline{\overline g}/\chi}$, $E\!=\!\eta/\big[2h(\xi)\overline{\overline g}\big]$ in Def. 2.6). \ep

\medskip

{\bf Remark 2.} We stress that the theorem holds also if
$\rho=\infty$. In the latter case $\xi$ is $\sigma_M$, if the latter
is finite, an arbitrary positive constant, if also
$\sigma_M=\infty$.

\medskip

Next, we are going to extend some of the previous results {\it in the large}.

\sect{Boundedness of the solutions and asymptotic stability in the large}
\label{largestability}

\begin{theorem}
Assume that: conditions (\ref{condi1}-\ref{condi3}), and possibly
either one of (\ref{conseq}'), are fulfilled with $\rho=\infty$
and $\tau< 1$; the function $g(t)$ defined by
(\ref{defgB}) fulfills  $\overline{\overline g}<\infty$; (\ref{CdotCondi1}) is fulfilled. Then:
\begin{enumerate}
\item the solutions of (\ref{eq}) are uniformly bounded;
\item the null solution of (\ref{eq}) is exponential-asymptotically stable in the large.

\hskip-.8truecm If only (\ref{CdotCondi2}), instead of (\ref{CdotCondi1}), is satisfied, then:
\item the solutions of (\ref{eq}) are eventually uniformly bounded;
\item the null solution of (\ref{eq}) is eventually exponential-asymptotically stable in the large.
\end{enumerate}
\label{thm2}
\end{theorem}

\bp~ As noted,
$r(\sigma)$ can be inverted to an increasing map
$r^{-1}:[0,r_M[\!\to\! [0,\sigma_M[$, whence also
\beq
\beta(\delta):=r^{-1}\left[\frac{\sqrt{\overline{\overline g}}
B(\delta)}{\sqrt{\chi}}\right] \label{defbeta}
\eeq
defines an increasing map
$\beta:[0,\delta_M[\!\to\! [0,\sigma_M[$, where
$\delta_M\!:=\!B^{-1}(r_M\sqrt{\chi}/\sqrt{\overline{\overline g}})$.
Note that $\beta(\delta)\!>\!\delta$.
An immediate consequence of (\ref{defbeta}) is
\beq
\frac{\overline{\overline g}B^2(\delta)}{\chi}=r^2[\beta(\delta)]=
\frac{\beta^2(\delta)}{1\!+\!\gamma_3[\beta(\delta)]}. \label{conseqbeta}
\eeq
From (\ref{Cdotip}) it immediately follows that
\beq
s(\delta):=\bar t\{\gamma_3[\beta(\delta)]\}  \: \left\{\ba{l}
=0 \qquad \mbox{ if (\ref{CdotCondi1}) is fulfilled,} \\[8pt]
<\infty \qquad \mbox{ if (\ref{CdotCondi2}) is fulfilled.}  \ea \right.
 \label{sdelta}
\eeq
We can
now show that for any $\delta\!\in]0,\delta_M[$, $t_0\!\ge\! s(\delta)$
\beq
d(t_0)<\delta\quad \qquad\Rightarrow \quad \qquad d(t) < \beta(\delta),
\qquad\forall t\ge t_0. \label{subtesi'}
\eeq
{\it Ad absurdum}, assume that there
exists a finite $t_2\!>\!t_0$ such that (\ref{subtesi'}) is fulfilled
for all $t\in[t_0,t_2[$, whereas
\beq
d(t_2)=\beta(\delta).                \label{absurdum'}
\eeq
The negativity of the rhs(\ref{Ineq1})
implies that $W(t)\equiv W\{u,u_t,t;\gamma_3[\beta(\delta)],\theta\}$
is a decreasing function of $t$ in $[t_0,t_2]$.
Using (\ref{Ineq2}),
(\ref{Ineq3}) and the (\ref{conseqbeta}) we find the following
contradiction with (\ref{absurdum'}):
$$
\chi d^2\!(t_2)\le W(t_2)< W(t_0)\le
\{1\!+\!\gamma_3[\beta(\delta)]\}g(t_0)B^2\left[d(t_0)\right]<
\{1\!+\!\gamma_3[\beta(\delta)]\}\overline{\overline g} B^2\!(\delta)=\chi
\beta^2(\delta).
$$

\medskip
Formula (\ref{subtesi'}) together with (\ref{sdelta}) proves
statements 1., 3.  under the assumption $\tau\!\in\![0,1[$, because
then by (\ref{defsigmaM}) $\delta_M=\infty$, so that we can choose
any $\delta\!>\!0$ in Definition 2.3.

With the above choice of $\theta$, by   (\ref{subtesi'}),
(\ref{Ineq3'})
 we find that for $t\!\ge\! t_0\!\ge\!  s(\delta)$ the Liapunov functional
$W_\delta(t)\equiv
W\big\{u,u_t,t;\gamma_3\big[\beta(\delta)\big],\theta(\delta)\big\}$
fulfills \beq W_\delta(t)\!\le \! h(\delta)\overline{\overline
g}d^2(t); \label{inter} \eeq this, together with (\ref{Ineq1})
implies $\dot W_\delta(t)\le -\eta
W_\delta(t)/[h(\delta)\overline{\overline g}]$ and (by means of the
comparison principle \cite{Yos66})
$W_\delta(t)<W_\delta(t_0)\exp\left[-\eta
(t\!-\!t_0)/[h(\delta)\overline{\overline g}]\right]$. From the
latter inequality, (\ref{Ineq2}) and  (\ref{inter}) with $t\!=\!t_0$
it follows
$$
d^2(t)\le \frac{W_\delta(t)}{\chi}<\frac{W_\delta(t_0)}{\chi}
\exp\left[-\frac{\eta}{h(\delta)\overline{\overline g}}
(t\!-\!t_0)\right] \le\frac{h(\delta)\overline{\overline g}}{\chi}
\exp\left[-\frac{\eta}{h(\delta)\overline{\overline g}}
(t\!-\!t_0)\right] \, d^2(t_0)
$$
for all $t\ge t_0\ge s(\delta) $. Recalling again (\ref{sdelta}), we
see that the latter formula proves statements 2., 4. \ep

In the case $\tau\!\ge\!1$ we find, by (\ref{defsigmaM}),
$$
\delta_M\!=\!B^{-1}\!\left(r_M\frac{\sqrt{\chi}}{\sqrt{\overline{\overline
g}}}\right)\!=\!B^{-1}\!\left\{\left[\frac{\tau\!-\!1}{1\!+\!\gamma_{31}}\right]^{\frac{\tau\!-\!1}{2\tau}}\!\frac{\sqrt{\chi}}{\sqrt{\overline{\overline
g}\tau\gamma_{32}^{1/{\tau}}}}\right\}.
$$
From the $\theta$-dependence of $\gamma_{31},\gamma_{32}$
[formulae (\ref{defthetagamma3}), (\ref{defgamma1})]
we see that $\delta_M$ decreases with $\theta$,
so we cannot exploit the freedom in the choice of $\theta$ to make
$\delta_M$  as large as we wish. This prevents us from
extending the results of the previous  theorem to the case $\tau\!\ge\!1$.

We can prove boundedness and asymptotic stability in the large even
for some unbounded $g(t)$, provided $\tau=0$.

\begin{theorem}
Assume that: conditions (\ref{condi1}-\ref{condi3}), and possibly
either one of (\ref{conseq}'), are fulfilled with $\rho=\infty$
and $\tau=0$; the function $g(t)$ defined by
(\ref{defgB}) fulfills (\ref{AsymStabCondi}); either (\ref{CdotCondi1})
or (\ref{CdotCondi2}) is fulfilled. Then:
\begin{enumerate}
\item the solutions of (\ref{eq}) are bounded;
\item the null solution of (\ref{eq}) is asymptotically stable in the large.
\end{enumerate}

\label{thm3}
\end{theorem}

\bp~ The condition $\tau=0$ means that $\gamma$ does not depend on $\sigma$;
then $r^{-1}(\beta)=\beta\sqrt{1\!+\!\gamma}$,
which is an increasing map $r^{-1}:I\!\to\! I$.
For any fixed $t_0$ setting
\beq
\tilde\beta(\alpha;t_0)\!:=\!r^{-1}\left[\frac{\sqrt{g(t_0)}B(\alpha)}{\sqrt{\chi}}\right]=B(\alpha)\frac{\sqrt{g(t_0)(1\!+\!\gamma)}}{\sqrt{\chi}}
\label{defbeta''}
\eeq
also defines an increasing map
$\tilde\beta:I\!\to\! I$, with $\tilde\beta(\alpha;t_0)\!>\!\alpha$.
We now prove statement 1., i.e. for any $\alpha\!>\!0$,
$t_0\!\ge\! \kappa\!:=\!\bar t(\gamma)$, 
\beq
d(t_0)<\alpha\qquad \qquad\Rightarrow
\qquad \qquad d(t) < \tilde\beta(\alpha;t_0)\qquad \forall t\!\ge\!t_0.\label{subtesi"}
\eeq
{\it Ad absurdum}, assume that
there exist a finite $t_2\!\in\![t_0,t]$ such that (\ref{subtesi"})
is fulfilled for all $t\in[t_0,t_2[$, whereas
\beq
d(t_2)=\tilde\beta(\alpha;t_0). \label{absurdum"}
\eeq
The
negativity of the rhs(\ref{Ineq1}) implies that $W(t)\equiv
W\{u(t),u_t(t),t;\gamma,\theta\}$ is a decreasing function of $t$ in
$[t_0,t_2]$. Using (\ref{Ineq2}), (\ref{Ineq3}) and
(\ref{defbeta''}) we find the following contradiction with
(\ref{absurdum"}):
$$\chi d^2\!(t_2)\le W(t_2)< W(t_0)\le
(1\!+\!\gamma)g(t_0)B^2\left[d(t_0)\right]<
(1\!+\!\gamma)g(t_0) B^2\!(\alpha)=\chi \tilde\beta^2(\alpha;t_0),
\qquad\qquad \mbox{Q.E.D.}
$$
By Theorem \ref{thm1} the null solution of (\ref{eq}) is stable.
Moreover, by (\ref{subtesi"}) relation (\ref{Ineq3}) becomes
$$
W(t)\le \tilde h(\alpha,t_0)g(t)d^2(t),  \qquad \qquad \qquad
\tilde h(\alpha,t_0):=\left(1\!+\!\gamma\right)
\left\{1\!+\!m\big[\tilde\beta(\alpha;t_0)\big]\right\}, 
$$
which,
together with (\ref{Ineq1}), implies $\dot W(t)\le -\eta W(t)/[\tilde h
g(t)]$ and (by means of the comparison principle \cite{Yos66})
$W(t)<W(t_0)\exp\left[-\eta \int^t_{t_0}dz/[\tilde hg(z)]\right]$, whence,
for all $t>t_0\ge\kappa$,
\bea
d^2(t) &\le& \frac{W(t)}{\chi}<
 \frac{W(t_0)}{\chi} \exp\!\left[-\frac \eta {\tilde h}\!\int\limits^t_{t_0}\!\!\frac{dz}{g(z)}\right]\le
\frac{\tilde hg(t_0)}{\chi}d^2(t_0) \exp\!\left[-\frac \eta {\tilde
h}\!\int\limits^t_{t_0}\!\!\frac{dz}{g(z)}\right]\nn[10pt]
&<&\frac{\tilde h(\alpha,t_0)g(t_0)}{\chi}\alpha^2
\exp\!\left[-\frac \eta {\tilde
h(\alpha,t_0)}\!\int\limits^t_{t_0}\!\!\frac{dz}{g(z)}\right]\nonumber
\eea The function $G(t)\!:=\!\int^t_{t_0}dz/g(z)$ is increasing and
by (\ref{AsymStabCondi}) diverges with $t$, what makes the rhs go to
zero as $t\to\infty$; more precisely, we can fulfill Def. 2.7
defining the corresponding function $T(\alpha,\nu,t_0,u_0,u_1)$ by
the condition that the rhs of the previous equation equals
$\nu_0^2\!:=\!\min\{\nu^2,\alpha^2\}$ at $t=t_0\!+\!T$, or equivalently that
$$
G(t_0\!+\!T)=-\frac {\tilde h(\alpha,t_0)}{\eta}\log\left[ \frac
{\chi\,\nu^2_0} {\tilde h(\alpha,t_0)\,\alpha^2}\right]
$$
(the rhs is positive as the argument of the logarithm is less than
1, by the definitions of $\chi,\tilde h$ and by the inequality
$\nu_0/\alpha\le 1$); this proves statement 2. \ep

\sect{Examples}
\label{examples}

Out of the many examples of forcing terms fulfilling (\ref{condi3})
we just mention $F(z)=b \sin(\omega z)$ (this has $F_z(z)\!\le\!
b\omega\!=:\!k$), which makes (\ref{eq}) into a modification of the
sine-Gordon equation, and the possibly non-analytic ones
$F(z)=-b|z|^qu$ with $b>0$, $q\ge 0$ (this has $F_z(z)\!\le\!
0\!=:\!k$), or $F(z)=b|z|^qu$ (this has
$F_z(z)\!=\!b(q\!+\!1)|z|^q\!<\! b(q\!+\!1)|\rho|^q\!=:\!k$ if
$|z|\!<\!\rho$). Out of the many examples of $t$-dependent
coefficients that fulfill (\ref{condi2}-\ref{condi1}) and either
(\ref{CdotCondi1}) or (\ref{CdotCondi2}), but not the hypotheses of
the theorems of \cite{DacDan98,DanFio00,DanFio05}, we just mention
the following ones:

\begin{enumerate}
\item $\varepsilon(t)=\varepsilon_0(1\!+\!t)^{-p }$
with constant $\varepsilon_0,p \!\ge\!0$ and $C\equiv
C_0\equiv$constant, with
$C_0\!>\!\frac{4(1\!+\!\varepsilon_0)k}{3\!+\!\varepsilon_0}
$. As a consequence $\overline{\varepsilon}\!=\!0\!\le\!\varepsilon
\!\le\!\varepsilon_0=\overline{\overline{\varepsilon}} $,
$\overline{\dot\varepsilon}\!=\!-\!p
\varepsilon_0\le\dot\varepsilon\!=\!-p \varepsilon_0[1\!+\!t]^{-\!p
\!-\!1} \!\le\!0\!=\!\overline{\overline{\dot\varepsilon}}$,
$\ddot\varepsilon\!=\!p (p \!+\!1)\varepsilon_0[1\!+\!t]^{-\!p
\!-\!2} \!\ge\!0\!=\!\overline{\ddot\varepsilon}$ [condition
(\ref{condi2})$_4$ is fulfilled],
($\varepsilon,\dot\varepsilon,\ddot\varepsilon\!\to\! 0$ as
$t\!\to\! \infty$). Conditions (\ref{condi2})$_1$-(\ref{condi2})$_3$
are fulfilled with $\mu\!=\!C/(1\!+\!\varepsilon_0)$. We find
$g(t)=C_0\!+\!p \varepsilon_0[1\!+\!t]^{-\!p \!-\!1}\!+\!1$, whence
$\overline{\overline g}=C_0\!+\!p \varepsilon_0\!+\!1$. Finally we
assume that $a'\!>\!0$ and $a$ fulfills (\ref{condi1})$_1$. Then
Theorems \ref{thm1},
 \ref{thm2}, apply: the null solution of (\ref{eq}) is
uniformly stable and uniformly  exponential-asymptotically
stable; it is also uniformly bounded and exponential-asymptotically
stable in the large if in addition $\rho=\infty$, $\tau\!<\!1$.

One can check that if we had adopted the same Liapunov functional as
in \cite{DanFio00,DanFio05} formulae (4.2), i.e. $W$ of (\ref{321})
with $\theta\!=\!0\!=\!a'$, for $p \!>\!1$ (namely
$\varepsilon\!\to\!0$ sufficiently fast as $t\!\to\! \infty$) we
would have not been able to prove the asymptotic stability.

\item
$\varepsilon(t)=\varepsilon_0(1\!+\!t)^p$, $C(t)=C_0(1\!+\!t)^q$,
with $1\ge q\ge p\ge 0$,  $\varepsilon_0\!\ge\!0$ and $C_0$
fulfilling
$$
C_0\!>\!p\varepsilon_0, \qquad\qquad
C_0\!>\!\frac{4(1\!+\!\varepsilon_0)k\!+\!2p\,\varepsilon_0}
{3\!+\!\varepsilon_0}.
$$
If $q,p\!>\!0$ then $C(t),\varepsilon(t)$ diverge as
$t\!\to\!\infty$. We immediately find
$\varepsilon(t)\!\ge\!\varepsilon_0\!=\!\overline{\varepsilon}$,
$\dot\varepsilon\!=\!p \varepsilon_0(1\!+\!t)^{p \!-\!1}\!\ge\! 0 $,
$\ddot\varepsilon\!=\!p (p \!-\!1)\varepsilon_0(1\!+\!t)^{p
\!-\!2}\!\le\! 0$, $\overline{\ddot\varepsilon}\!=\!p (p
\!-\!1)\varepsilon_0$ [condition (\ref{condi2})$_4$ is fulfilled],
$C(t)\!\ge\!C_0$,
$$
\frac{C\!-\!\dot\varepsilon}{1\!+\!\varepsilon}=\frac{C_0(1\!+\!t)^q
\!-\!p \varepsilon_0(1\!+\!t)^{\!p\!-\!1}}
{1\!+\!\varepsilon_0(1\!+\!t)^p}= \frac{ C_0(1\!+\!t)^{\!q \!-\!p}
\!-\!p{\varepsilon_0}(1\!+\!t)^{-1}}{(1\!+\!t)^{\!-\!p}\!+\!\varepsilon_0}
 \ge  \frac{ C_0
\!-\!p{\varepsilon_0}}{1\!+\!\varepsilon_0},
$$
and conditions (\ref{condi2})$_1$-(\ref{condi2})$_3$ are fulfilled
with $\mu=(C_0\!-\!p\varepsilon_0)/(1\!+\!\varepsilon_0)$. Moreover,
$\dot C=qC_0(1\!+\!t)^{\!q \!-\!1}\to 0$ as  $t\!\to\! \infty$
[condition (\ref{CdotCondi2}) is fulfilled];  $g(t)$ grows as $t^q$,
implying that (\ref{AsymStabCondi}) is fulfilled. Finally we assume
that $a$ fulfills (\ref{condi1})$_1$ [condition (\ref{condi1})$_2$
is already satisfied] . Then Theorem \ref{thm1} applies: the null
solution of (\ref{eq}) is asymptotically stable. If in addition
$\rho=\infty$, $\tau\!=\!0$ then Theorem \ref{thm3} applies, and the
null solution is also bounded and asymptotically stable in the
large.

\item $\varepsilon(t)$ fulfilling
$\overline{\overline{\varepsilon}}\!<\!\infty$,
$\overline{\overline{\dot\varepsilon}}\!<\!\infty$,
$\overline{\dot\varepsilon}\!>\!-\infty$,
$\overline{\ddot\varepsilon}\!>\!-\infty$ [condition
(\ref{CdotCondi2})]; we note that this includes
periodic $\varepsilon(t)$.  $C(t)=C_0\!+\! C_1(1\!+\!t)^{-q}$ with
constant $C_0,C_1,q$ fulfilling $C_1>0$, $q\ge 0$ and
$$
C_0\!>\!\max\left\{0,\overline{\overline{\dot\varepsilon}},
\frac{4(1\!+\!\overline{\overline{\varepsilon}})k\!+\!2\overline{\overline{\dot\varepsilon}}}
{3\!+\!\overline{\overline{\varepsilon}}}\right\}, \qquad\qquad
C_0\!\ge\!k.
$$
Then  conditions (\ref{condi2})$_1$-(\ref{condi2})$_3$ are fulfilled
with
$\mu=(C_0\!-\!\overline{\overline{\dot\varepsilon}})/(1\!+\!\overline{\overline{\varepsilon}})$.
Moreover, $\dot C\!\le\! 0$ [condition (\ref{CdotCondi1}) is
fulfilled]. We find $g(t)\le C_0\!+\!C_1\!-\!\overline{\dot\varepsilon}\!+\!1\!=:\!\overline{\overline
g}\!<\!\infty$. Finally we assume that $a'\!>\!0$ and $a$ fulfills
(\ref{condi1})$_1$. Then Theorems \ref{thm1},
 \ref{thm2}, apply: the null solution of (\ref{eq}) is
uniformly stable and uniformly  exponential-asymptotically
stable. It is also uniformly bounded and exponential-asymptotically
stable in the large if in addition $\rho=\infty$, $\tau\!<\!1$.

\end{enumerate}

\end{document}